\documentclass{article}
\usepackage{amssymb,amsmath,amscd, amsthm,stmaryrd,verbatim, mathrsfs}

\newtheorem{Th}{Theorem}[section]

\newtheorem{rmk}{Remark}

\newcommand{\bb}{\mathbb}
\newcommand{\ms}{\mathscr}
\newcommand{\mr}{\mathrm}

\usepackage{hyperref}
\begin{document}

\title{Dirac and Yang Monopoles Revisited}
\author{Guowu Meng\\
\small{\it Department of Mathematics, Hong Kong Univ. of Sci. and Tech.}\\
\small{\it Clear Water Bay, Kowloon, Hong Kong}\\
\small{Email: mameng@ust.hk} }
\date{May 29, 2007}
\maketitle
\begin{abstract}
The Dirac monopoles in $3$-space and their generalization by C. N.
Yang to $5$-space are observed to be just the Levi-Civita spin
connections of the cylindrical Riemannian metric on the $3$- and
$5$- dimensional punctured spaces respectively. Their
straightforward generalization to higher dimensions is also
investigated.
\end{abstract}



\section{Introduction}
The main purpose of this note is to provide a geometric description
of the Dirac and Yang monopoles from which a generalization to
higher dimensions follows naturally. We believe that it is good to
be aware of this observation for anyone who is interested in
monopoles and their related objects in theoretical physics. For
example, by using these generalized Dirac monopoles, the MICZ-Kepler
problems has been extended \cite{meng05} beyond dimension five.

\medskip
In order to understand the background about our general result, it
is helpful to have a quick review of the motivation and the main
result of Ref. \cite{yang78}.

In Ref. \cite{yang78}, Yang starts with the following observation:
the Dirac monopole with magnetic charge $g$ is uniquely determined
by the following two properties:

(a) For any closed surface $\Sigma$ around the origin of ${\bb
R}^3$, the magnetic flux through $\Sigma$ is $4\pi g$ ($g$ is a
nonzero half integer), i.e., equivalently,
\begin{eqnarray}\int_\Sigma {1\over 2\pi} F=2g.
\end{eqnarray}
(Here $F$ is the field strength and is singular only at the origin
of ${\bb R}^3$.)

(b) It is $\mr {SO}(3)$ symmetric.

\medskip
Next he starts to search for a generalization of the Dirac
monopole on ${\bb R}^3\setminus\{0\}$ to an $\mr {SU}(2)$-gauge
field on ${\bb R}^5\setminus\{0\}$, i.e., an $\mr {SU}(2)$-gauge
field on ${\bb R}^5\setminus\{0\}$ satisfying

($\mr a^\prime$) For any closed hyper-surface $\Sigma$ around the
origin of ${\bb R}^5$,
\begin{eqnarray}\int_\Sigma \mr {Tr}F^2\neq 0,
\end{eqnarray}
(Here $F$ is the field strength and is singular only at the origin
of ${\bb R}^5$; and $\mr {Tr}$ is the trace in the defining
representation of $\mr {SU}(2)$.)

($\mr b^\prime$) It is $\mr {SO}(5)$ symmetric.

\noindent Here is Yang's main result in Ref. \cite{yang78}: There
are two and only two solutions $\alpha$ and $\beta$ satisfying ($\mr
a^\prime$) and ($\mr b^\prime$) and are characterized by
\begin{eqnarray}\int_\Sigma {1\over 8\pi^2}\mr {Tr} F^2=1
\quad\hbox{and}\quad \int_\Sigma {1\over 8\pi^2}\mr {Tr} F^2=-1
\end{eqnarray}respectively.

In Ref. \cite{yang78}, the local formula for vector potentials of
the monopoles are written down explicitly, but no conceptual reason
is given. However, the two characterization properties of Yang
monopoles force us to believe that there must be a clean conceptual
description of the monopoles in terms of geometry, and that belief
turns out to be true.

\subsection{Main Results} In terms of spherical
coordinates, the flat metric on ${\bb R}^{2n+1}$ is
\begin{equation}
ds_E^2 = dr^2+r^2d\Omega^2
\end{equation}
where $r$ is the radius and $d\Omega^2$ is the round metric for
the unit sphere. On ${\bb R}^{2n+1}\setminus\{0\}$, the flat
metric is conformally equivalent to the {\bf cylindrical metric}:
\begin{equation}
ds^2 = {1\over r^2}ds_E^2={1\over r^2}dr^2+d\Omega^2.
\end{equation}
Note that $\mr{SO}(2n+1)$ is clearly a group of isometry on the
Riemannian manifold $({\bb R}^{2n+1}\setminus\{0\}, ds^2)$. Note
also that the gamma matrix associated to the unit radial
directional vector of ${\bb R}^{2n+1}\setminus\{0\}$ splits the
spin bundle on $({\bb R}^{2n+1}\setminus\{0\}, ds^2)$ into two sub
vector bundles of rank $2^{n-1}$, with respect to this splitting,
the Levi-Civita spin connection splits into $A_+$ and $A_-$ on the
respective sub vector bundles. Note also that $\mr{Spin}(2n+1)$ is
a group of symmetry of the plus (or minus) spin bundle which
leaves $A_+$ (or $A_-$) invariant.

We are now ready to state
\begin{Th}[Main Theorem]\label{main}
Let $(A_+, A_-)$ be the
Levi-Civita spin connection on the Riemannian manifold $({\bb
R}^{2n+1}\setminus\{0\}, ds^2)$, then
\begin{enumerate}

\item  $A_\pm$ are $\mr {U}(2^{n -1})$-gauge fields on ${\bb
R}^{2n+1}\setminus\{0\}$, in fact $\mr {SU}(2^{n -1})$-gauge
fields on ${\bb R}^{2n+1}\setminus\{0\}$ if $n>1$.

\item Let $F_\pm$ be the gauge field strength respectively,
$\Sigma$ a closed hyper-surface around the origin of ${\bb
R}^{2n+1}$, then
\begin{eqnarray}\int_\Sigma {1\over n!}\mr {Tr}(-{F_\pm\over 2\pi})^{n}=\pm 1
\end{eqnarray}
respectively, where $\mr {Tr}$ is the trace in the defining
representation of ${\mr U}(2^{n-1})$.

 \item When $n=1$, $A_\pm$ are respectively the Dirac monopole with
fundamental plus or minus magnetic charge $g=\pm{1\over 2}$.

\item When $n =2$, $A_\pm$ are respectively the Yang monopoles
$\alpha$ and $\beta$ introduced in Ref. \cite{yang78}.

\end{enumerate}
\end{Th}

\begin{rmk}
One can show that, when $n>1$, there are no other
$\mr{SU}(2^{n-1})$-gauge field on $({\bb R}^{2n+1}\setminus\{0\},
ds^2)$ such that 1) it is $\mr{Spin}(2n+1)$-symmetric, 2)
$\int_\Sigma \mr {Tr}(-{F_\pm\over 2\pi})^{n}$ is nontrivial for
the field strength $F$ and closed hyper-surface $\Sigma$ around
the origin of ${\bb R}^{2n+1}$. But this is less interesting, so
its proof will be omitted.
\end{rmk}

\underline{Added in May, 2007}. The results presented here were
posted in the archive as math-ph/0409051 in September, 2004. I was
told by I. Cotaescu that he wrote Ref. \cite{Cotaescu05} in the
winter 2004/2005 in which virtually the same results were obtained.
While we had different motivations, we reached the same destination
independently.

\section{Proof of Theorem \ref{main}}

Here we assume the readers are familiar with the basic facts on
Clifford algebra and spin groups, a good reference is Ref.
\cite{ML}. We assume $e_1$, \ldots, $e_k$ are the standard basis of
$\bb R^k$ and if $x$, $y$ are in $\bb R^k$ then $xy$ is the
multiplication of $x$ with $y$ in Clifford algebra. Note that
$xy+yx=-2x\cdot y$ where $x\cdot y$ is the dot product of $x$ with
$y$. We also assume the reader is familiar with the definition of
connection on a principal bundle in the sense of Ehresmann (see page
358 of Ref. \cite{Spivak}).

In our recent paper \cite{meng03}, we observe that there is a
canonical principal $\mr {Spin}(2n)$-bundle over $\mr S^{2n}$:
\begin{eqnarray}
\mr {Spin}(2n)\to\mr{Spin}(2n+1)\to \mr S^{2n} \end{eqnarray} with
a canonical $\mr{Spin}(2n+1)$-symmetric $\mr{Spin}(2n)$-connection
\begin{eqnarray}\omega(g)=\mr {Pr}(g^{-1}dg)\end{eqnarray}
where $\mr {Pr}$ is the orthogonal projection onto the Lie algebra
of $\mr {Spin}(2n)$. It is clear that $\omega(hg)=\omega(g)$ for
any $h\in\mr {Spin}(2n+1)$. Note that the bundle and the
connection here are precisely the principle spin bundle and the
Levi-Civita connection of $\mr{S}^{2n}$ equipped with the standard
round metric.

As a Riemannian manifold,  $({\bb R}^{2n+1}\setminus\{0\},
ds^2)=\bb R\times \mr{S}^{2n}$ where $\bb R$ is standard real line
with the standard flat metric and $\mr{S}^{2n}$ is the standard
unit sphere with the standard round metric. Let
$$p_2:\hskip 5pt \bb R\times \mr{S}^{2n}\to \mr{S}^{2n}$$
be the projection map, then it is clear that the spin bundle on
$({\bb R}^{2n+1}\setminus\{0\}, ds^2)$ is just the pullback by
$p_2$ of the spin bundle on $\mr{S}^{2n}$, so is the spin
connection.

Note that $\mr{Spin}(2n)$ has two fundamental spin representations
$\rho_\pm$ of dimension $2^{n-1}$, so we have two associated
vector bundles over $\bb R^{2n+1}\setminus\{0\}$ of rank
$2^{n-1}$, hence two associated canonical
$\mr{Spin}(2n+1)$-symmetric $\mr{U}(2^{n-1})$-connections $A_\pm$
--- the ones we mentioned in the introduction.

We are ready to present the proof of our main theorem.

\begin{description}

\item[Step1]: To get explicit expressions for the canonical $\mr
{Spin}(2n+1)$-symmetric connection $\omega(g)=\mr {Pr}(g^{-1}dg)$,
we need to choose an open cover (a family of open sets of $\bb
{R}^{2n+1}\setminus\{0\}$ whose union is equal to $\bb
{R}^{2n+1}\setminus\{0\}$) and a gauge over each of the open sets
of the open cover, i.e., a smooth section of the principal bundle
over each of the open sets of the open cover.

The open cover chosen here consists of two open sets: $V_\pm$,
where $V_\pm$ is $\bb R^{2n+1}$ with the positive/negative
$(2n+1)$-st axis removed, i.e., $\bb R\times \mr{S}_\pm$ where
$\mr{S}_\pm$ is the sphere with north/south pole removed.

 A point on $V_-$ can be written as $x=r\cos \theta
e_{2n+1}+r\sin \theta y$ where $0\le \theta <\pi$, $y\in \mr
S^{2n-1}$ and $r>0$. On $V_-$, we choose this gauge:
\begin{eqnarray}
g(x)=\cos{\theta\over 2}-\sin{\theta\over 2}ye_{2n+1}.
\end{eqnarray}
(it is easy to see that $x=g(x)e_{2n+1}g(x)^{-1}$, so $g$ is
indeed a smooth section over $V_-$.) Then, under this choice of
gauge, we can calculate the gauge potential $\ms
A(x):=-i\omega(g(x))=-i\mr {Pr}(g^{-1}(x)dg(x))$ and get
\begin{eqnarray}\label{potential}
\fbox{$\ms A(x)=i(\sin{\theta\over 2})^2y\,dy$}\,;
\end{eqnarray}
we can also calculate the gauge field strength $\ms F=d\ms A+i\ms
A^2$ and get
\begin{eqnarray}\label{field}
\ms F(x)={i\over 2}\sin\theta\,d\theta\, y\,dy+{i\over
4}(\sin{\theta})^2\,dy\,dy.
\end{eqnarray}

A point on $V_+$ can be written as $x=r\cos \theta e_{2n+1}+r\sin
\theta y$ where $0< \theta \le\pi$, $y\in \mr S^{2n-1}$ and $r>0$.
On $V_+$, we choose this gauge:
\begin{eqnarray}
\tilde g(x)=\left(\cos{\theta\over 2}-\sin{\theta\over
2}ye_{2n+1}\right)ye_{2n}=\cos{\theta\over
2}ye_{2n}-\sin{\theta\over 2}e_{2n+1}e_{2n}.
\end{eqnarray}
Under this choice of gauge, we have
\begin{eqnarray}\label{potential'}
\fbox{$\tilde {\ms A}(x)=-i(\cos{\theta\over
2})^2e_{2n}y\,dye_{2n}$}\,
\end{eqnarray}
and
\begin{eqnarray}\label{field'}
\tilde {\ms F}(x)=e_{2n}y\, {\ms F}(x)\, ye_{2n}={i\over
2}\sin\theta\,d\theta\,e_{2n} y\,dye_{2n}-{i\over
4}(\sin{\theta})^2\,e_{2n}dy\,dye_{2n}.
\end{eqnarray}

Let $A_\pm =\rho_\pm(\ms A)$ and $F_\pm=\rho_\pm(\ms F)$. The only
nontrivial topological number we are calculating here is
\begin{eqnarray}
c_\pm:=\int_{\mr S^{2n}}{1\over n!}\mr {Tr}\left(-{F_\pm\over
2\pi} \right)^n=\int_{\mr S^{2n}}{1\over n!}\mr
{Tr}_{\rho_\pm}\left(-{\ms F\over 2\pi} \right)^n=\int_{\mr
S^{2n}}{1\over n!}\mr {Tr}_{\rho_\pm}\left(-{\tilde {\ms F}\over
2\pi} \right)^n,
\end{eqnarray}
where $\mr{S}^{2n}$ should be viewed as the hyper-surface $1\times
\mr{S}^{2n}$ in $\bb R\times \mr{S}^{2n}$.

And the calculation of $c_\pm$ is not hard:
\begin{eqnarray}
c_\pm&=&{2\over (n-1)!}\left(-{i\over 8\pi}\right)^n\int_{\mr
S^{2n}}(\sin\theta)^{2n-1}\, d\theta \mr
{Tr}_{\rho_\pm}\left(y\,dy (dy\, dy)^{n-1} \right)\cr &= &{2\over
(n-1)!}\left(-{i\over 8\pi}\right)^n\int_0^\pi
(\sin\theta)^{2n-1}\, d\theta \int_{\mr S^{2n-1}} \mr
{Tr}_{\rho_\pm}\left(y\,dy (dy\, dy)^{n-1} \right)\cr &=&\pm
{n(2n-1)!!\over (2\pi)^n}\mr{vol}(\mr B^{2n})\int_0^\pi
(\sin\theta)^{2n-1}\, d\theta\cr &=&\pm 1.
\end{eqnarray}
Here $\mr{vol}(\mr B^{2n})$ is the volume of the unit $2n$-ball
and we assume that the orientation on spheres is the standard one
and the convention that
\begin{eqnarray}
\rho_\pm\left(e_1\cdots e_{2n}\right)=\pm i^n I_{2^{n-1}}
\end{eqnarray}
is adopted, here $I_{2^{n-1}}$ is the identity matrix of order
$2^{n-1}$.

In the above calculation, if we replace $\mr{S}^{2n}$ by any
closed hyper-surface $\Sigma$ around the origin of ${\bb
R}^{2n+1}$, the result is still the same, that is because of the
Stokes theorem and the Bianchi identity. So point 2 of the main
theorem is proved.

\item[Step2]: Note that (\ref{potential}) and (\ref{potential'})
are the explicit formulae for the gauge potentials, and
(\ref{field}) and (\ref{field'}) are the explicit formulae for the
corresponding gauge field strengths. When $n>1$, we claim that
$\rho_\pm(\ms A)$ and $\rho_\pm(\tilde {\ms A})$ are
$\mr{SU}(2^{n-1})$-gauge potentials, that is because
$\mr{Tr}_{\rho_\pm}(e_ie_j)=0$ if $i\neq j$: For example,
\begin{eqnarray}
\mr{Tr}_{\rho_\pm}(e_1e_2)&=&\mr{Tr}_{\rho}\left(e_1e_2{1\over
2}(1\pm (-i)^ne_1\cdots e_{2n})\right) \cr & =&{1\over
2}\left(\mr{Tr}_{\rho}(e_1e_2)\pm(-i)^{n}\mr{Tr}_{\rho}(e_3e_4\cdots
e_{2n})\right)\cr & =&{1\over
2}\left(\mr{Tr}_{\rho}(e_2e_1)\pm(-i)^{n}\mr{Tr}_{\rho}(e_{2n}e_3e_4\cdots
e_{2n-1})\right)\cr &=& -{1\over 2}\left(\mr{Tr}_{\rho}(e_1e_2)\pm
(-1)^{n-1}\mr{Tr}_{\rho}(e_3e_4\cdots e_{2n})\right)\cr &=&0.
\end{eqnarray}
So point 1 of the main theorem is proved.

\item[Step3]: To prove points 3 and 4, we can do the explicit
calculations and then compare with explicit formulae in Ref.
\cite{yang78}. A quick way to prove them is to use the two
characterization properties of the Dirac or Yang monopoles that we
mentioned in the introduction.
\end{description}

\begin{thebibliography}{10}

\bibitem{meng05} G. W. Meng: ``MICZ-Kepler
problems in all dimensions", {\em J. Math. Phys.} {\bf 48} (2007),
032105.

\bibitem{yang78}
C. N. Yang: ``Generalization of Dirac's monopole to $SU_2$ gauge
fields", {\em J. Math. Phys.} {\bf 19} (1978) 320-328.

\bibitem{Cotaescu05}
I. I. Cotaescu: ``Generalized Dirac monopoles in non-Abelian
Kaluza-Klein theories", {\em Nucl. Phys. B} 719 (2005) 140-164.


\bibitem{ML}
H. B. Lawson and M.L. Michelsohn: {\em Spin geometry}, Princeton
Mathematical Series, {\bf 38}, Princeton University Press,
Princeton, NJ, 1989.

\bibitem{Spivak}
M. Spivak: {\em A comprehensive introduction to differential
geometry}, Vol. II, Second edition, Publish or Perish, Inc.,
Berkeley, 1979.

\bibitem{meng03} G. W. Meng: ``Geometric construction of the Quantum Hall Effect in all even dimensions", {\em J. Phys. A: Math. Gen.} {\bf 36} (2003) 9415-9423.

\end{thebibliography}
\end{document}